\documentstyle[12pt,epsfig]{article}
\let\section=\subsection     \let\subsection=\subsubsection 
 
\font\schwell=schwell at 12pt
\renewcommand{\l}{\schwell L\/}

\begin{document}
\begin{center}
   {\large \bf MODELING $\mbox{\boldmath$J/\psi$}$ PROPERTIES}\\[2mm]
   {\large \bf AT FINITE TEMPERATURE}\\[5mm]
   K.~HAGLIN$^{\,a}$ and
   C.~GALE$^{\,b}$\\[5mm] 
   {$^{a}$\small \it  Department of Physics, Astronomy and Engineering 
   Science\\
   Saint Cloud State University \\
   720 Fourth Avenue South, St. Cloud, MN 56301 \ USA\\[8mm]}
   {$^{b}$\small \it  Physics Department\\
   McGill University \\
   3600 University Street\\
   Montr\'eal, QC, H3A 2T8 \ Canada\\[8mm]} 
\end{center}

\begin{abstract}\noindent
   We model {\it J\/}/$\psi$ using an effective lagrangian approach and 
   calculate its spectral function in a gas of light mesons at finite
   temperature.  We explore the hadronic landscape estimating cross sections
   for elastic and inelastic channels.  Effects of form factors are tested
   in a general, but consistently gauge invariant manner.  Relevance to 
   heavy ion experiments is discussed.
\end{abstract}

\section{Introduction}
Experimental efforts at Brookhaven's Relativistic Heavy Ion
Collider will soon bring to fruition laboratory studies
of nuclear systems heated and compressed to energy densities far exceeding
that of the proton.  Quantum chromodynamics (QCD), which is believed 
be the appropriate description in the subatomic domain, predicts that matter 
produced in this regime could have its subhadronic degrees of freedom 
liberated from hadronic boundaries and allowed to move about as a plasma of 
quarks and gluons\cite{me96}.  Identification of the quark-gluon plasma 
(QGP) concerns much of the current activities as models and interpretation 
seem to point toward a common conception regarding the nature of the 
transition or crossover from ordinary hadronic matter to QGP.   However, many 
issues remain unsettled\cite{qm97} providing a wealth of opportunities 
in the vibrant and rapidly advancing research field of
ultra-relativistic heavy ion physics.

Possibly the best available tool for studying nuclear systems in such states
of excitation are electromagnetic probes of photons and lepton pairs.
Each has its kinematical advantages, but both share the property that
once produced in the system, undeflected flight to detectors ensues.
Since virtual photons couple directly to the neutral
component of the vector hadronic current, dileptons have been raised
to a level of premier importance as they allow study of in-medium
properties of the vector mesons.   In the low mass sector, effects of the
$\rho$ meson clearly appear in proton-nucleus reactions
whereas in heavy-ion experiments a significant broadening of the
``$\rho\,$'' distribution is observed\cite{ceres}.   The importance of 
broadening 
effects through relatively high scattering rates were highlighted several
years ago in Ref.~\cite{kh95}.  Since then sophisticated hadronic models 
have been developed which seek to include in a consistent formalism these 
and other effects into vector meson spectral functions\cite{rr96}.  The 
emerging lore favors a $\rho$ distribution in matter which is broadened 
essentially beyond recognition\cite{rhoinmedium}. 

In the higher mass sector, $J/\psi$ appears as a promising tool for 
spectroscopy.  Within conventional hadronic scenarios it is expected to 
contribute a measurable muon-pair signal, whereas in QGP scenarios color
screening effectively inhibits $c\bar{c}$ binding and limits $J/\psi$ 
survival probabilities\cite{ms86}.  $J/\psi$ yields are
therefore expected to be severely suppressed when plasma is produced.
The notion has recently gained a great deal of attention as $J/\psi$ 
yields from Pb+Pb reactions at 158 $A\/$GeV\cite{jpsiexpt} show 
``anomalous'' suppression compared simple extrapolations from lighter
projectiles' results.  Models of absorption on hadronic comovers
are able to provide interpretation for the lighter system\cite{jpsitheory},
but not consistently for the lead results.  Very intriguing analyses 
involving QGP scenarios have been put forward\cite{dk98} which are able 
to explain the yields.  Meanwhile, purely hadronic approaches have been 
proposed too, but necessary input of $J/\psi$ scattering cross sections 
with light hadrons\cite{jpsiplush} are up to now quite 
uncertain\cite{kr99}.  We report here on results of effective
field theoretical methods of estimating cross sections for 
$J/\psi$ with light hadrons and we use them to construct a
spectral function for $J/\psi$ in hot hadronic matter.

\section{Effective Lagrangian}

Hadronic interactions are here modeled with meson exchange, consequently 
a symmetry embodying strangeness and charm is needed.  We therefore begin
with {\it SU\/}(4) and introduce pseudoscalar ($\phi$ = $\varphi_{a}
\lambda_{a}$) and vector ({\it V\/}$^{\mu}$ = ${v\/}_{a}^{\mu}\lambda_{a}$)
meson matrices, where $\varphi_{a}$ and ${v\/}_{a}^{\mu}$ are pseudoscalar
and vector multiplets and the $\lambda$s are {\it SU\/}(4) 
generators.  The large charm quark mass and its symmetry breaking
effects are duly noted, but since we use physical mass eigenstates 
and we incorporate empirical constraints on the model where possible,
we expect reasonably reliable results as evidenced by the coupling
constants' near universality\cite{kh00}.

The free meson Lagrangian is written as 
\begin{eqnarray}
\mbox{\l}_0 \ = \ {\rm Tr}(\partial^\mu \phi^\dag \partial_\mu \phi)\ - \
{\rm Tr}\left( \left(\partial_\mu V_\nu^\dag\right)
\left(\partial^\mu V^\nu - \partial^\nu V^\mu \right) \right)\ + \ {\rm
mass\ terms}\ .
\end{eqnarray}
The pseudoscalar meson mass matrix that leads to properly normalized 
mass terms is 
\begin{eqnarray}
\phi\ = \ \left(
\begin{array}{cccc}
\frac{\pi^0}{\sqrt{2}} + \frac{\eta}{\sqrt{6}} + \frac{\eta_c}{\sqrt{12}}
& \pi^+ & K^+ & \bar{D}^0 \\
\pi^- & -\frac{\pi^0}{\sqrt{2}} + \frac{\eta}{\sqrt{6}} +
\frac{\eta_c}{\sqrt{12}} & K^0 & D^- \\
K^- & \bar{K}^0 & - \eta \sqrt{\frac{2}{3}} + \frac{\eta_c}{\sqrt{12}}
& D_s^- \\
D^0 & D^+ & D_s^+ & - 3 \frac{\eta_c}{\sqrt{12}}  \\ 
\end{array}
\right) \ ,
\end{eqnarray}
while that for the vector multiplet is (suppressing the Lorentz index)
\begin{eqnarray}
V\ = \ \left(
\begin{array}{cccc}
\frac{\rho^0}{\sqrt{2}} + \frac{\omega}{\sqrt{6}} + \frac{J/\psi}{\sqrt{12}}
& \rho^+ & {K^*}^+ & \bar{{D^*}}^0 \\
\rho^- & -\frac{\rho^0}{\sqrt{2}} + \frac{\omega}{\sqrt{6}} +
\frac{J/\psi}{\sqrt{12}} & {K^*}^0 & {D^*}^- \\
{K^*}^- & \bar{{K^*}}^0 & - \omega \sqrt{\frac{2}{3}} + \frac{J/\psi}{\sqrt{12}}
& {D^*}_s^- \\
{D^*}^0 & {D^*}^+ & {D^*}_s^+ & - 3 \frac{J/\psi}{\sqrt{12}}  \\ 
\end{array}
\right) \ .
\end{eqnarray}

We then introduce interactions through a gauge covariant minimal substitution
$\partial_{\mu}\mbox{\it f\/}\rightarrow \mbox{\it D\/}_{\mu}\mbox{\it 
f\/\rm =} \partial_{\mu}\mbox{\it f\/\rm + [{\it A\/}}_{\mu}\mbox{\rm ,\it 
f\,\rm{]}}$, where  $\mbox{\it A\/}_{\mu}$ = -$ig/2\,V_{\mu}$.  As an
aside remark, we note that the appearance  of the $g/2$ instead of
a mere $g$ is nothing but a choice of convention.  Since the model is
calibrated to data, the choice is irrelevant.
We must collect terms up to order $g^{\,2}$ for a consistently gauge invariant
description for the hadronic currents.  They are (since $\phi^{\dag}$
= $\phi$ and {\it V\/}$^{\dag}$ = {\it V\/})
\begin{eqnarray}
\mbox{\l}_{\rm\,int} &=& ig\,\mbox{\rm Tr}(\phi{\it 
V\/}^{\mu}
\partial_{\mu}\phi-\partial^{\mu}\phi{\it V\/}_{\mu}
\phi) 
+\frac{1}{2}\,g^{\,2}\,\mbox{\rm Tr}(\phi{\it V\/}^{\mu}{\it 
V\/}_{\mu}\phi-\phi{\it V\/}^{\mu} 
\phi{\it V\/}_{\mu}) \nonumber\\
&+&ig\mbox{\rm Tr}\left(\partial^{\mu}{\it V\/}^{\nu\,}
\lbrack
{\it V\/}_{\mu}, {\it V\/}_{\nu}
\rbrack + \lbrack
{\it V\/}^{\mu\,}, {\it V\/}^{\nu\,}
\rbrack 
\partial_{\mu}{\it V\/}_{\nu}\right)
+{g^{\,2}}\mbox{\rm Tr}\left(
{\it V\/}^{\mu\,}
{\it V\/}^{\nu\,}
\left\lbrack{\it V\/}_{\mu},{\it V\/}_{\nu}\right\rbrack\right).\ \ \
\end{eqnarray}

Carrying out the matrix algebra, we arrive at the totality of interaction
terms allowed in the symmetry group.  

The model is calibrated to observed hadronic decays,
tuned to respect vector dominance, or as a last resort, to respect
{\it SU\/}(4) symmetry.  Details can be found in Ref.~\cite{kh00}.

\subsection{The $\mbox{\boldmath{$J/\psi+h\/$}}$ cross section}
Of particular importance are cross sections for a light hadron
to knock apart the charmonium leaving {\it D\/},
{\it D\/}$^{*}$ or antiparticles when appropriate for conservation of 
relevant quantities.  We present in Fig.~1 the results for the 
$\pi$-, {\it K\/}-, and $\rho$-induced dissociation.
\begin{figure}[h]
\vspace*{7.5cm}
\hspace*{6cm}
\includegraphics{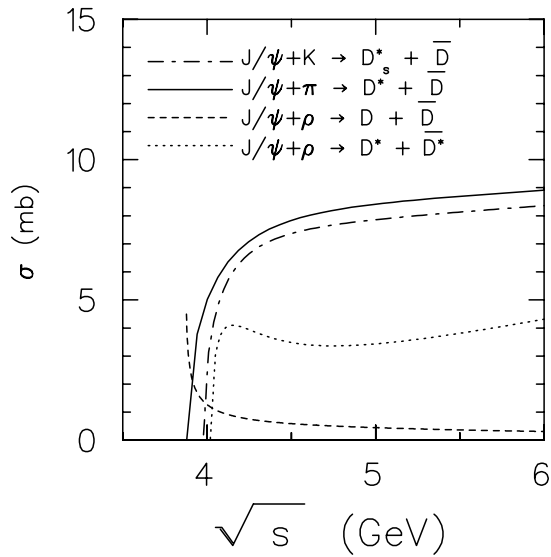}
\caption{\small{Dissociation cross sections for pions, kaons and rho mesons.
Bands of uncertainties (not shown) are estimated to be $\sim$ 20--50\%.}}
\end{figure}
Purely elastic $J/\psi+h\rightarrow J/\psi+h$ are found to be
of order femtobarns for pions, microbarns for rho and nanobarns for kaons, 
clearly too small for any significance.

\section{The $\mbox{\boldmath{$J/$}}\mbox{\boldmath$\psi$}$ spectral function}

We next place $J/\psi$ in a strongly interacting medium and explore 
possible modifications in spectral properties as compared to the vacuum.
From previous studies on vector mesons, one has seen that two-loop 
effects are much stronger than one-loop\cite{kh95,cgjk}.  We have checked
to see this trend continue for charmonium.  We here adopt the approach
to neglect one-loop effects and use free masses throughout.

We calculate the expected broadening of the $J/\psi$ distribution 
from collisions with light pseudoscalar and vector mesons in the hot
hadronic matter. The extra width induced in the distribution by a 
reaction of type $J/\psi \, 2 \rightarrow 3 4$, where 2, 3, and 4 are 
arbitrary species is \cite{kh95,more_width_calcs}. 

\begin{eqnarray}
\Gamma (\omega, \vec{p} ) & = & \frac{1}{2 \omega} \int \, d\Omega\, 
n_2 (E_2) (1 + n_3
(E_3)) (1 + n_4(E_4))| \overline{ {\cal M} (J/\psi 2 \rightarrow 3 
4)}|^2\ , \ \
\label{width}
\end{eqnarray}
where $\omega\ = \ \sqrt{\vec{p\,}^2 + m_{J/\psi}^2}$, $\vec{p}$ being
the three-vector of the $J/\psi$.  Note that 3 or 4 can be a $J/\psi$. 
In Eq. (\ref{width}), 
\begin{eqnarray}
d\Omega\ &=& \ d\bar{p}_2 d\bar{p}_3 d\bar{p}_4 (2 \pi)^4 \delta ( p + p_2
- p_3 - p_4 ),  
\end{eqnarray}
where we have used shorthand notation
\begin{eqnarray}
d\bar{p}_i\ &=& \ \frac{d^3 p_i}{(2\pi )^3\, 2 E_i}\ .
\end{eqnarray}

A direct connection can be drawn between the rate in Eq.~(\ref{width}) and 
the eventual structure of the spectral function, utilizing at the intermediate
stages such field theoretical concepts as the $J/\psi$ propagator and 
imaginary part of the self-energy\cite{more_width_calcs}.
In an on-shell approximation, one has for the full spectral function
\begin{eqnarray}
A_{J/\psi} (\omega, \vec{p\,})\ = \ \frac{2\, m_{J/\psi}\, 
\Gamma_{J/\psi}}{(p^2 - m_{J/\psi}^2)^2 + m_{J/\psi}^2
\Gamma_{J/\psi}^2 }\ ,
\label{spectralfunction}
\end{eqnarray}
where $\Gamma_{J/\psi}$ contains the vacuum width as well as 
contributions from elastic and inelastic collisions.  Discussion
of Eq.~(\ref{spectralfunction}) in the context of more
general formalism will be published elsewhere\cite{khcg00}.

Now we consider $J/\psi$ in a finite temperature gas consisting 
of $\pi$s, $K$s, $\rho$s and $K^{*}$s. The
spectral function at 150 MeV is shown below in 
Fig.~2.  Severe broadening of the
spectral distribution with an accompanying suppression of the peak
is immediately clear already for temperatures on the order of the
pion mass.   Studies of temperature dependence as well as
specific channels' contributions will also be included
in Ref.~\cite{khcg00}.
\begin{figure}[h]
\vspace*{7.5cm}
\hspace*{6cm}
\includegraphics{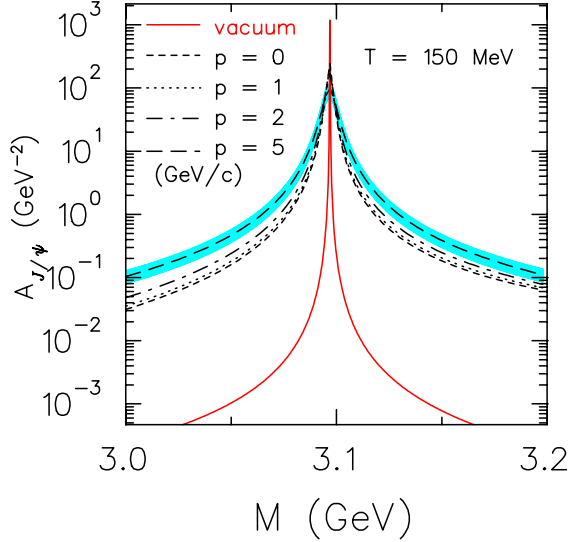}
\caption{\small{Spectral function for $J/\psi$ at 150 MeV temperature.
The medium is comprised of light mesons up to 1 GeV mass.
A (typical) band of uncertainty appears over the 5 GeV results.}}
\end{figure}

\section{Effects of form factors}

Up to now pointlike hadrons have been considered.  Effective 
lagrangian approaches are not considered complete until
finite size effects are incorporated, typically 
accomplished with vertex form factors.  Here we outline the procedure
for implementation.  Each $t$-channel Feynman graph is given a product
of monopoles, one for each vertex, having the following structure
\begin{eqnarray}
h(t) &=& \left({\Lambda_{1}+t_{max}-m_{\alpha}^{2}\over
\Lambda_{1}+t_{max}-t}\right)\cdot
\left({\Lambda_{2}+t_{max}-m_{\alpha}^{2}\over
\Lambda_{2}+t_{max}-t}\right)\ ,
\end{eqnarray}
where $\Lambda_{1}$ and $\Lambda_{2}$ are cutoff parameters depending
on the species of on-shell particle, and $m_{\alpha}$ is the mass of 
the exchanged meson.  Two cases will be examined: first we choose
a fixed cutoff for all vertices of 2, 3 and 4 GeV, and second we allow
each vertex to be given its own $\Lambda_{i}$ = $m_{i}$, where
$m_{i}$ is the mass of the on-shell charmed meson entering or
leaving the vertex.

Exchange graphs, or $u$-channels, are given similar form-factor 
structure.  Let us call the function $g(u)$.  Finally, contact
graphs are given general Lorentz structure with several expansion coefficients.
The idea is then to determine the coefficients (in terms of Lorentz
invariants and $h$ and $g$) which result in a manifestly
gauge invariant amplitude for the most general case of functions $h$
and $g$.  Due to space limitation, we do not include here the full
expression.  More complete details will be published elsewhere\cite{khcg00}.

As an example of form-factor effects, we show in Fig.~3 the dissociation 
cross section for one of the pion channels ($K$ and $\rho$ cross
sections are similarly affected).  We are inclined to point
to the results for $\Lambda_{\alpha} = m_{\alpha}$ as likely being
the most consistent, although means of constraining the form factors are
under investigation\cite{khcg00}.
\begin{figure}[h]
\vspace*{7.5cm}
\hspace*{6cm}
\includegraphics{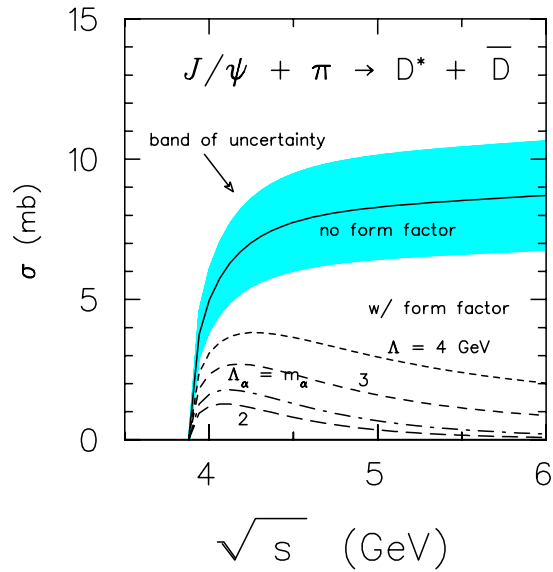}
\caption{\small{Dissociation cross section for $J/\psi+\pi\rightarrow
\bar{D}+D^{*}$ using a range of form factor parameters.  See
discussion in the text for more detail.}}
\end{figure}
\section{Conclusions}

We have studied the spectral function for $J/\psi$ modified from
its vacuum structure due to interactions with a gas of light mesons.
Results suggest that the spectral function gets considerably modified due
mostly to dissociation-type reactions.  An environment much like
the one we studied here is known to occur in ultrarelativistic heavy ion 
collisions.  We stress that the approach we have taken is fully consistent
in terms of conserving currents and respecting gauge invariance.
Some issues will require further attention before we
can take the next step of employing a spacetime model to ultimately
compare with heavy ion data.  Since the $SU\/$(4) symmetry is known to be 
only approximately valid, some coupling strengths are not completely
pinned down.  Also, the final results will depend fairly sensitively 
on what we do with the form factors.   Work on these and related
issues is in progress\cite{khcg00}.

\vskip 0.5 \baselineskip
\leftline{\large\bf Acknowledgment}
\vskip 0.5 \baselineskip
\noindent K.H. is supported by the National Science Foundation
under Grant No. 9814247, and C.G. is supported by the
Natural Science and Engineering Research Council of Canada and by the
Fonds FCAR of the Quebec Government.

\end{document}